\documentclass[11pt]{article}
\usepackage{amssymb}
\usepackage{amsmath}
\usepackage[dvips]{graphicx} 
\begin{document}

\begin{center}

{\Large Classification of dispersion equations for homogeneous 
dielectric--magnetic uniaxial materials}\\
\vskip 1cm

{\bf Ricardo A. Depine\footnote{E-mail: rdep@df.uba.ar}$^{(a,b)}$, 
Marina E. Inchaussandague\footnote{Corresponding Author. E-mail: mei@df.uba.ar}$^{(a,b)}$ 
and Akhlesh  Lakhtakia\footnote{E-mail: akhlesh@psu.edu}$^{(c)}$} \\
\bigskip

{$^{(a)}$ GEA~---~Grupo de Electromagnetismo Aplicado, Departamento de F\'{\i}sica,}\\
{Facultad de Ciencias Exactas y Naturales, Universidad de Buenos Aires,}\\
{Ciudad Universitaria, Pabell\'{o}n I, 1428 Buenos Aires, Argentina}\\
{$^{(b)}$ CONICET~---~Consejo Nacional de Investigaciones Cient\'{\i}ficas y T\'ecnicas,}\\
{Rivadavia 1917, Buenos Aires, Argentina}\\
{$^{(c)}$ CATMAS~---~Computational and Theoretical Materials Sciences Group,\\
Department of Engineering Science and
Mechanics,\\ Pennsylvania State University, University Park, PA
16802--6812, USA}\\

\end{center}

\bigskip
\noindent ABSTRACT\\
\baselineskip 0.7 cm
The geometric representation at a fixed frequency of the wavevector 
(or dispersion) surface $\omega(\vec k)$ for lossless, homogeneous  
dielectric--magnetic uniaxial materials is explored, when
 the elements of the relative permittivity and permeability tensors 
of the material can have any sign.
Electromagnetic plane waves propagating inside the material can exhibit 
dispersion surfaces in the form of  ellipsoids of revolution,
hyperboloids of one sheet, or hyperboloids of two sheets. 
Furthermore, depending on the relative orientation 
of the optic axis, the intersections of these surfaces
with fixed planes of propagation can
be circles, ellipses, hyperbolas, or straight lines. The obtained
understanding is used to study the reflection and refraction  of 
electromagnetic plane waves due to a planar interface with an isotropic medium.


\bigskip

\section{Introduction}\label{sec:intro}
\baselineskip 0.7 cm
Recent developments in mesoscopic (i.e., structured but effectively homogeneous)
materials have significantly broadened 
the range of available electromagnetic constitutive properties, thereby allowing 
the realization of solutions to Maxwell's equations which 
could have been previously regarded as mere academic exercises. 
Materials having effectively   negative
real  permittivity 
and permeability have been constructed \cite{shelbyscience,parazzoliprl90,houckprl90}
from arrays of conducting wires \cite{wires} and arrays of split ring resonators 
\cite{splitring}. Such composite materials~---~often called  
metamaterials~---~exhibit a negative index of refraction
in certain frequency regimes \cite{LMW02}. 
Under these conditions, the phase velocity vector is in the opposite direction 
of the energy flux, for which reason they have been called 
\emph{negative--phase--velocity} (NPV) materials \cite{LMW03,Boardman}. 

NPV metamaterials synthesized thus far are actually anisotropic 
in nature, and any hypothesis about their isotropic behavior holds 
only under some restrictions on propagation direction and polarization 
state.  In  anisotropic NPV
materials, the directions of power flow and phase velocity are not 
necessarily antiparallel but~---~more generally~---~have a negative 
projection of one on the other \cite{ML_PRE}. 
Since the use of anisotropic NPV materials offers flexibility 
in design and ease of fabrication, attention has begun to be drawn to such materials 
\cite{hu-chui,LS03,smith_prl90,smith_josab21}.

Natural crystals are characterized by permittivity and permeability tensors 
with the real part of all their elements positive, a fact that leads to dispersion 
equations in the form of closed surfaces. On the other hand, a relevant 
characteristic of NPV metamaterials is that the real parts of the
elements of their permittivity 
and permeability tensors can have different signs in different frequency ranges. 
As an example, Parazzoli \emph{et al.} \cite{parazzoliprl90} 
demonstrated negative refraction using $s$--polarized microwaves and samples 
for which the permittivity and permeability tensors have certain eigenvalues 
that are negative real.
Under such circumstances, dispersion equations 
are topologically similar to 
\emph{open} surfaces \cite{eritsyan}. 
Consequently,  the intersection of
a dispersion surface and a fixed plane of propagation may be a curve of
an unusual 
shape, compared with its analogs for natural crystals. For example, extraordinary plane 
waves in a simple dielectric (nonmagnetic) uniaxial medium can exhibit dispersion 
curves which are hyperbolic, instead of the usual elliptic curves characteristic 
of natural uniaxial crystals \cite{eritsyan,LMDhyper}. 
In recent studies on the characteristics of anisotropic materials with hyperbolic 
dispersion curves, new phenomenons have been identified, such as omnidirectional 
reflection~---~either from a single boundary \cite{hu-chui} or from multilayers 
\cite{Liu}~---~and the possibility of an infinite number of refraction channels 
due to a periodically corrugated surface \cite{GEM_NJP,GEM_scalar}. 

In this paper, we are interested in studying the conditions under which the combination 
of permittivity and permeability tensors with the real parts of their 
elements of arbitrary sign, leads to closed or open dispersion surfaces for a homogeneous 
dielectric--magnetic uniaxial material. To characterize this kind of material, 
four constitutive scalars are needed: 
\begin{itemize}
\item $\epsilon_{\parallel}$ and $\mu_{\parallel}$, 
which are the respective elements of the relative permittivity and relative 
permeability tensors along the optic axis; and 
\item $\epsilon_{\perp}$ and $\mu_{\perp}$, 
which are the elements of the two tensors in the plane perpendicular to the optic axis.
\end{itemize}
These scalars have positive real parts for natural crystals, but their real parts
can have any sign for artificial (but still effectively homogeneous) materials. 
The dispersion equation for plane waves in such a material can be factorized 
into two terms, leading to the conclusion that the material supports the 
propagation of two different types of linearly polarized waves, called magnetic 
and electric modes \cite{LVV91ije,LVV91jmo}. 

The relative permittivity and permeability 
tensors,  $\tilde\epsilon$ and $\tilde\mu$, are real symmetric when
dissipation can be ignored. Then, each tensor can be classified as \cite{Lut}: 
\begin{itemize}
\item[(i)] positive definite, if all eigenvalues are positive; 
\item[(ii)] negative definite, if all eigenvalues are negative; and 
\item[(iii)] indefinite, if it has both negative and positive eigenvalues.
\end{itemize}
Thus, the relative permittivity tensor is positive definite if $\epsilon_\perp >0$ 
and $\epsilon_\parallel >0$; it is negative definite if $\epsilon_\perp <0$ 
and $\epsilon_\parallel <0$; and it is indefinite if 
$\epsilon_\perp  \epsilon_\parallel <0$. 
In the present context, we exclude constitutive tensors with null eigenvalues. 
A similar classification applies to the relative permeability tensor. 
If both $\tilde\epsilon$ and $\tilde\mu$ are positive 
definite, the material is of the positive--phase--velocity (PPV) kind. 

The plan of this paper is as follows. Considering the different possible 
combinations of  $\tilde\epsilon$ and $\tilde\mu$, we show in Section 
\ref{sec:disp-surf} that magnetic and electric propagating modes can exhibit 
dispersion surfaces which are
\begin{itemize}
\item[(a)] ellipsoids of revolution, 
\item[(b)] hyperboloids of one sheet, or 
\item[(c)] hyperboloids of two sheets. 
\end{itemize}
As a byproduct of our analysis, we also obtain different possible combinations 
of $\tilde\epsilon$ and $\tilde\mu$ that preclude the propagation 
of a mode~---~either electric, magnetic or both~---~inside the material. 
In Section \ref{sec:disp-curves} we study the intersection between the dispersion 
surfaces and a fixed plane of propagation that is arbitrarily oriented with respect 
to the optic axis. We show that, depending on the relative orientation of 
the optic axis, different dispersion curves, in the form of circles, ellipses, 
hyperbolas or even straight lines, can be obtained. Previous studies on dielectric--magnetic materials 
with indefinite constitutive tensors only considered planes of propagation 
coinciding with coordinate planes, thus failing to identify the 
singular case of linear dispersion equations. These results are used in 
Section \ref{sec:boundary} to discuss the reflection and refraction 
of electromagnetic plane waves due to a planar interface between a dielectric--magnetic 
uniaxial material and an isotropic medium. Concluding remarks are provided 
in Section \ref{sec:conc}. 
An $\exp(-i\omega t)$ time--dependence is implicit, with $\omega$ as angular 
frequency, $t$ as time, and $i=\sqrt{-1}$.

\section{Dispersion surfaces}\label{sec:disp-surf}

The relative permeability and permittivity tensors of the anisotropic medium share the same optic axis denoted by the unit vector $\hat{c}$, and their four eigenvalues are denoted by $\epsilon_{\perp,\parallel}$ and  $\mu_{\perp,\parallel}$. In dyadic notation \cite{Chen}
\begin{equation}
\left.\begin{array}{l}
\tilde \epsilon =\epsilon_{\perp} \tilde I +
(\epsilon_{\parallel}-\epsilon_{\perp}) \;\hat c \;\hat c \\[5pt]
\tilde \mu =\mu_{\perp} \tilde I + (\mu_{\parallel}
-\mu_{\perp}) \;\hat c\;\hat c
\end{array}\right\}\,, \label{constit}
\end{equation}
with $\tilde I$ the identity dyadic. In this medium, two distinct plane 
waves can propagate in any given direction: 
\begin{itemize}
\item[(i)] electric modes, with dispersion equation
\begin{eqnarray}
\vec k \cdot\tilde \epsilon \cdot \vec k= 
k_0^2\, \mu_\perp \epsilon_\perp \epsilon_\parallel\,, \label{ke1}
\end{eqnarray}
and 
\item[(ii)] magnetic modes, with dispersion equation
\begin{eqnarray}
\vec k \cdot\tilde \mu \cdot \vec k = k_0^2 \,
\mu_\perp  \epsilon_\perp \mu_\parallel\,. \label{km1}
\end{eqnarray}
\end{itemize}
Here $\vec k$ is the wavevector and $k_0$ denotes the free--space wavenumber. 

We decompose the wavevector $\vec k=\vec k_{\parallel}+\vec k_{\perp}$ 
into its components parallel ($\vec k_{\parallel}$) and perpendicular 
($\vec k_{\perp}$) to the optic axis. After taking into account that 
\begin{eqnarray}
\vec k \cdot\tilde \epsilon \cdot \vec k = 
\epsilon_\perp \,(\vec k \times \hat c)^2 + 
\epsilon_\parallel\,(\vec k \cdot \hat c)^2 , \label{quade1}
\end{eqnarray}
(\ref{ke1}) for electric modes can be rewritten as
\begin{eqnarray}
\frac{k_{\perp}^2}{\epsilon_\parallel} + 
\frac{k_{\parallel}^2}{\epsilon_\perp} = k_0^2\, \mu_\perp\,. \label{ke2}
\end{eqnarray}
Analogously, (\ref{km1}) for magnetic modes can be expressed as
\begin{eqnarray}
\frac{k_{\perp}^2}{\mu_\parallel} + 
\frac{k_{\parallel}^2}{\mu_\perp} = k_0^2\, \epsilon_\perp\,. \label{km2}
\end{eqnarray}
Equations (\ref{ke2}) and (\ref{km2}) have both the  quadric form
\begin{eqnarray}
\frac{k_{\perp}^2}{A} + \frac{k_{\parallel}^2}{B} = 1\,,\label{common1}
\end{eqnarray}
which displays symmetry of revolution about the $k_\parallel$ axis 
in three--dimensional $\vec{k}$--space. The parameters $A$ and $B$ depend on the kind of mode (electric or 
magnetic) and their values determine the propagating or evanescent character of 
each mode and the geometric nature of the dispersion surface for propagating 
modes. 

\begin{figure}[ht] 
\begin{center} 
\begin{tabular}{c}
\includegraphics[width=16.4cm]{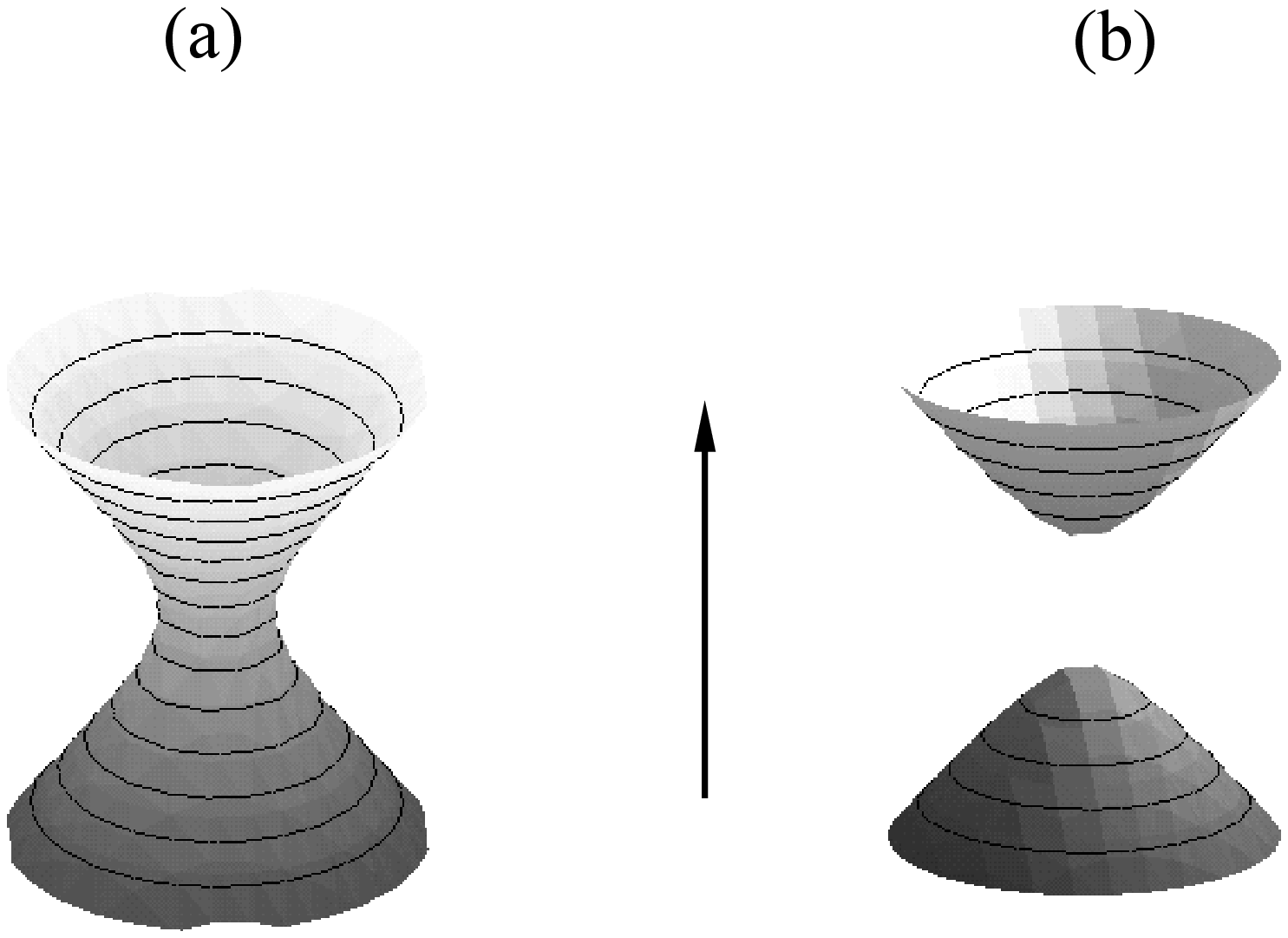} 
\end{tabular}
\end{center} 
\caption[example]{ \label{hyper12} Geometrical representations of
 (\ref{common1}). 
(a) $A>0$ and $B<0$, hyperboloid of one sheet; 
(b) $A<0$ and $B>0$: hyperboloid of two sheets.}
\end{figure}

One of the following conditions applies for a specific 
mode:
\begin{itemize}
\item[(i)] $A>0$ and $B>0$: the dispersion surface is an ellipsoid of revolution; 
\item[(ii)] $A>0$ and $B<0$: the dispersion surface is an hyperboloid of one sheet 
(Figure \ref{hyper12}a); 
\item[(iii)] $A<0$ and $B>0$: the dispersion surface is an hyperboloid of two sheets
(Figure \ref{hyper12}b); 
\item[(iv)] $A<0$ and $B<0$: the mode is evanescent. 
\end{itemize}
Depending on the particular combination of $\tilde\epsilon$ and $\tilde\mu$, 
we obtain from these conditions different dispersion surfaces. 
For example, the dispersion equations for electric and magnetic modes in natural
crystals are 
both represented by eq. \ref{common1} with $A>0$ and $B>0$, a fact that leads 
to the known result that electric and magnetic modes have dispersion surfaces 
in the form of either prolate or oblate ellipsoids of revolution. 
The same result is obtained for metamaterials with both 
constitutive tensors negative definite. 
When the analysis is repeated for all possible combinations between the 
four constitutive scalars 
$\epsilon_{\perp}$, $\epsilon_{\parallel}$, $\mu_{\perp}$ and $\mu_{\parallel}$, 
the  results summarized in Table 1 are obtained.

\section{Intersection with a fixed plane of propagation}\label{sec:disp-curves}

In the previous section, by considering plane wave propagation in an unbounded medium, 
we found the various geometric forms of the dispersion surfaces. 
At a specularly flat interface between two half-spaces
filled with linear homogeneous materials,  the tangential 
components of the wavevectors of the incident, transmitted and reflected plane waves 
must all be equal, and consequently, they all must lie in the same plane that is
orthogonal to the interface. This plane is the \emph{plane of propagation}.
Let us now investigate the 
kinds of dispersion curves obtained when  dispersion surfaces of the kind
identified in 
Section \ref{sec:disp-surf} intersect by a specific plane of propagation, 
arbitrarily oriented with respect to the optic axis $\hat c$. 

Without loss of generality, let the $xy$ plane
be the fixed plane of propagation in a cartesian coordinate system; furthermore, let
$\hat c= c_x \hat x + c_y \hat y + c_z \hat z$ and
$\vec k= k_x \hat x + k_y \hat y$. The 
dispersion equation (\ref{ke1}), for electric modes, can then be rewritten as the 
quadratic equation
\begin{eqnarray}
M_{11}\,k_x^2 + 2 \, M_{12}\, k_x\,k_y + M_{22}\, k_y^2 = F \;, \label{quade2}
\end{eqnarray}
where
\begin{equation}
\left.\begin{array}{l} 
M_{11} = \epsilon_{\perp} + (\epsilon_{\parallel}-\epsilon_{\perp}) c_x^2 \;\\[4pt]
M_{12} = (\epsilon_{\parallel}-\epsilon_{\perp})  c_x \,c_y \;\\ [4pt]
M_{22} = \epsilon_{\perp} + (\epsilon_{\parallel}-\epsilon_{\perp}) c_y^2 \;\\[4pt]
F      = k_0^2\,\epsilon_{\parallel} \epsilon_{\perp}\,\mu_\perp
\end{array}\right\}\,. \label{qcoefic}
\end{equation} 
The dispersion equation (\ref{km1}) for magnetic modes also has the same
quadratic form, but now the coefficients $M_{11}$, $M_{12}$, $M_{22}$, and 
$F$ are obtained by the interchange 
$\left\{ \epsilon_\parallel \longleftrightarrow \mu_\parallel,
\epsilon_\perp \longleftrightarrow \mu_\perp \right\}$ 
in (\ref{qcoefic}). 

The symmetric matrix 
\begin{equation}
\tilde M = \left[
\begin{array}{cc}
M_{11} & M_{12}\\
M_{12} & M_{22}
\end{array}\right]\,,
\end{equation} corresponding to the quadratic equation 
(\ref{quade2}) is defined by its three elements. 
This matrix can be diagonalized by rotating the $xy$ plane about the
$z$ axis by a certain angle, thereby eliminating the $k_x k_y$
term in (\ref{quade2}). With 
$\hat v_1$ and $\hat v_2$ denoting the orthonormalized eigenvectors of the
 matrix $\tilde M$, we can write $\vec k= k_1 \hat v_1 + k_2 \hat v_2$.
 Likewise, with 
 \begin{equation}
\left.\begin{array}{l} 
\lambda_{1}=\epsilon_\perp + (\epsilon_{\parallel}-\epsilon_{\perp}) (c_x^2+ c_y^2)\;\\
\lambda_{2}=\epsilon_\perp 
\end{array}\right\}\,. \label{eigen1}
\end{equation} 
denoting the eigenvalues of $\tilde M$, we get the dispersion curve
\begin{eqnarray}
\lambda_1\,k_1^2 + \lambda_2\,k_2^2 = F \; \label{quade3}
\end{eqnarray}
in the plane of propagation.

The dispersion curves
for the mode represented by (\ref{quade3}) can be classified by analyzing the signs of $\lambda_1$, $\lambda_2$ and $F$. In particular,
\begin{itemize}
\item[(i)] 
if $\lambda_1$, $\lambda_2$ and $F$
all  have the same sign, then the dispersion curve 
 in the fixed plane of propagation 
is an ellipse, with semiaxes along the directions $\hat v_1$ and $\hat v_2$; 
\item[(ii)] 
if $\lambda_1$ and $\lambda_2$ have both the same sign, but $F$ has the opposite sign, 
then the mode represented by (\ref{quade3}) is of the evanescent kind; 
\item[(iii)] 
if $\lambda_1$ and $\lambda_2$ have opposite signs, then the dispersion curve 
is a hyperbola, with semiaxes along the directions $\hat v_1$ and $\hat v_2$; 
\item [(iv)]
if one eigenvalue is equal to zero and the other (nonzero) eigenvalue has 
the same sign as $F$, then the dispersion curve is a straight line, 
parallel to the eigenvector associated with the null eigenvalue. 
\end{itemize} 
 
\section{Illustrative numerical results and discussion}\label{sec:boundary}

To illustrate the different possibilities for the dispersion curves,
let us present numerical results for the following two cases:
\begin{itemize}
\item[]{\it Case I:}  $\epsilon_{\perp}=-2.1$, $\epsilon_{\parallel}=1.9$, 
$\mu_{\perp}=1.3$ and $\mu_{\parallel}=-1.6$;	
\item[]{\it Case II:} 
$\epsilon_{\perp}=2.1$, $\epsilon_{\parallel}=-1.9$, $\mu_{\perp}=-1.3$ 
and $\mu_{\parallel}=1.6$. 
\end{itemize}
Both constitutive tensors thus are chosen to be indefinite.
According to
Table 1, the electric and magnetic modes
for both Case I and Case II have dispersion surfaces 
in the form of one--sheet hyperboloids of revolution, whose intersections 
with fixed planes of propagation are circles, ellipses, hyperbolas or straight 
lines~---~depending on the orientation of $\hat{c}$. 

Furthermore, to show the 
usefulness of our analysis in visualizing dispersion curves for boundary value 
problems, let us now consider that the anisotropic medium is illuminated by a plane wave 
from a vacuous half--space, the plane of incidence being the $xy$ plane. 
In terms of  (a) the angle $\theta_c$ between the optic 
axis and the $y$ axis and (b) the angle $\varphi_c$ between the $x$ axis 
and the projection of the optic axis onto the $xy$ plane, 
the   optic axis can be stated as
\begin{equation}
\hat{c}
=\hat{x}\sin \theta_c \,\cos \varphi_c +\hat{y}
\cos \theta_c + \hat{z}\sin \theta_c\,\sin \varphi_c\,\,, 
\label{opticaxis}
\end{equation}
and the eigenvalues $\lambda_j^E$, corresponding to electric modes 
can be written as
\begin{equation}
\left.\begin{array}{l} 
\lambda_1^E=\epsilon_\perp + (\epsilon_{\parallel}-\epsilon_{\perp}) 
(1-\sin^2 \theta_c \,\sin^2 \varphi_c)\,\\
\lambda_2^E=\epsilon_\perp 
\end{array}\right\}\,. \label{lambdaelec}
\end{equation}

For Case I, $F^E<0$, $\lambda_{2}^E=\epsilon_\perp<0$, 
whereas the sign of $\lambda_{1}^E$ depends on the optic axis 
orientation. From (\ref{lambdaelec}) we conclude for the electric modes
as follows:
\begin{itemize}
\item 
$\lambda_{1}^E>0$ if 
\begin{equation}
\sin^2 \theta_c \,\sin^2 \varphi_c < \frac{\epsilon_{\parallel}}{\epsilon_{\parallel}-\epsilon_{\perp}}\,\,,\label{E1p2n}
\end{equation}
and the dispersion curves are 
hyperbolas with semiaxes along the directions $\hat v_1^E$ and $\hat v_2^E$; 
\item 
$\lambda_{1}^E=0$ if 
\begin{equation}
\sin^2 \theta_c \,\sin^2 \varphi_c = \frac{\epsilon_{\parallel}}{\epsilon_{\parallel}-\epsilon_{\perp}}\,\,,\label{E102n}
\end{equation}
and the dispersion curves are
straight lines parallel to the direction associated with the 
eigenvector $\hat v_1^E$; and
\item 
$\lambda_{1}^E<0$
if  
\begin{equation}
\sin^2 \theta_c \,\sin^2 \varphi_c > \frac{\epsilon_{\parallel}}{\epsilon_{\parallel}-\epsilon_{\perp}}\,\,,\label{E1p2p}
\end{equation}
and the dispersion curves are 
ellipses with semiaxes along the directions of the eigenvectors 
$\hat v_1^E$ and $\hat v_2^E$. 
\end{itemize}
The same conclusions hold for electric modes in Case II. 

Analogously, the eigenvalues $\lambda_j^M$, corresponding to magnetic modes 
are as follows:
\begin{equation}
\left.\begin{array}{l} 
\lambda_1^M=\mu_\perp + (\mu_{\parallel}-\mu_{\perp}) 
(1-\sin^2 \theta_c \,\sin^2 \varphi_c)\\
\lambda_2^M=\mu_\perp
\end{array}\right\}\,, \label{lambdamagn}
\end{equation} 
For Case I, $F^M>0$ and $\lambda_{2}^M=\mu_\perp>0$.
From (\ref{lambdamagn}) we deduce that
\begin{itemize}
\item 
$\lambda_{1}^M<0$
if   
\begin{equation}
\sin^2 \theta_c \,\sin^2 \varphi_c < \frac{\mu_{\parallel}}{\mu_{\parallel}-\mu_{\perp}}\,\,,\label{M1p2n}
\end{equation}
and the dispersion curves   are 
hyperbolas with semiaxes along the directions $\hat v_1^M$ and $\hat v_2^M$; 
\item 
$\lambda_{1}^M=0$
if 
\begin{equation}
\sin^2 \theta_c \,\sin^2 \varphi_c = \frac{\mu_{\parallel}}{\mu_{\parallel}-\mu_{\perp}}\,\,,\label{M1p20}
\end{equation}
and the dispersion curves   are 
straight lines parallel to the direction associated of the 
eigenvector $\hat v_1^M$; 
\item 
$\lambda_{1}^M>0$
if  
\begin{equation}
\sin^2 \theta_c \,\sin^2 \varphi_c > \frac{\mu_{\parallel}}{\mu_{\parallel}-\mu_{\perp}}\,\,,\label{M1p2p}
\end{equation}
and the dispersion curves are
ellipses with semiaxes along the directions of the eigenvectors 
$\hat v_1^M$ and $\hat v_2^M$. 
\end{itemize}
The same conclusions hold for magnetic modes in Case II. 

Let $\varphi_c>0^\circ$ so that the optic axis is not wholly contained in the plane of incidence. There exist critical values of
$\theta_c$ at which the dispersion curve change from hyperbolic/elliptic to elliptic/hyperbolic.
By virtue of (\ref{E102n}), the critical value for electric modes
 is given by 
\begin{eqnarray}
\label{aaa1}
\sin \theta_c^E=\left[\frac{\epsilon_\parallel}
{(\epsilon_\parallel-\epsilon_\perp)\sin^2\varphi_c}\right]^{1/2}\,.
\end{eqnarray}
Likewise, the critical value
\begin{eqnarray}
\label{aaa2}
\sin \theta_c^M=\left[\frac{\mu_\parallel}
{(\mu_\parallel-\mu_\perp)\sin^2\varphi_c}\right]^{1/2}\,
\end{eqnarray}
for magnetic modes emerges from (\ref{M1p20}). 
Expressions (\ref{aaa1}) and (\ref{aaa2}) are valid for
both Cases I and II. 
At a critical value of $\theta_c$, the dispersion curve for the corresponding
mode is a straight line.

Suppose $\varphi_c=60^\circ$, so that  $\theta_c^E=52.73^{\circ}$ and $\theta_c^M=59.06^{\circ}$.
Then, for  $\theta_c=\theta_c^E$ the dispersion curves 
in the plane of incidence are straight lines (electric modes) and hyperbolas 
(magnetic modes); whereas for $\theta_c=\theta_c^M$, the dispersion curves are 
ellipses (electric modes) and straight lines (magnetic modes).

\begin{figure}[hbt] 
\begin{center} 
\begin{tabular}{c}
\includegraphics[width=7.5cm]{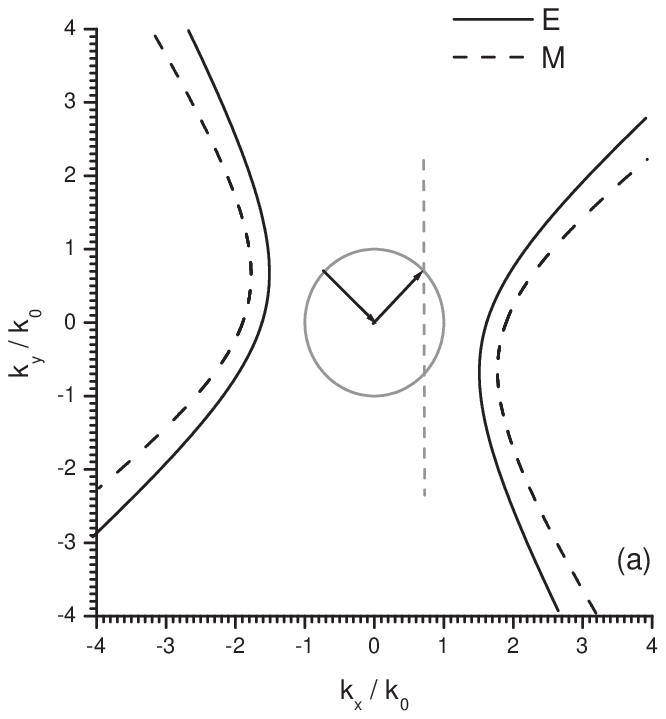} \hspace{0.3cm} 
\includegraphics[width=7.5cm]{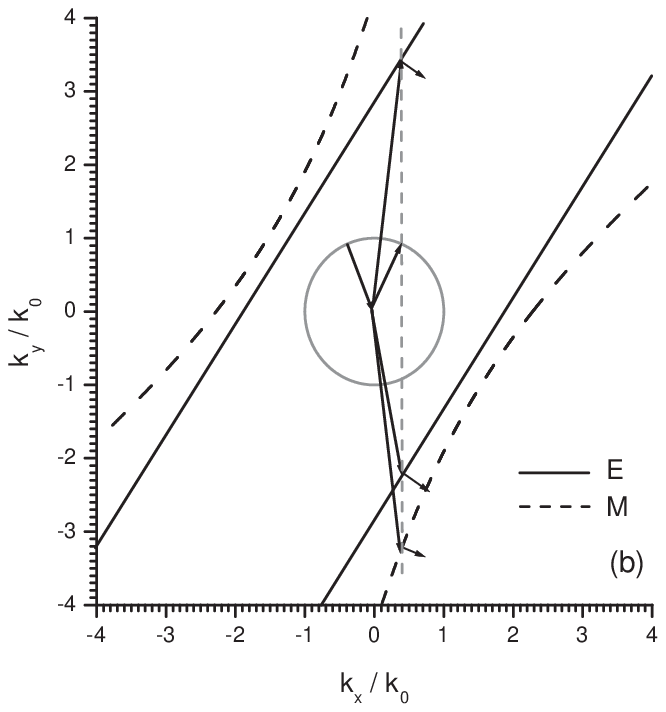}\\
\includegraphics[width=7.5cm]{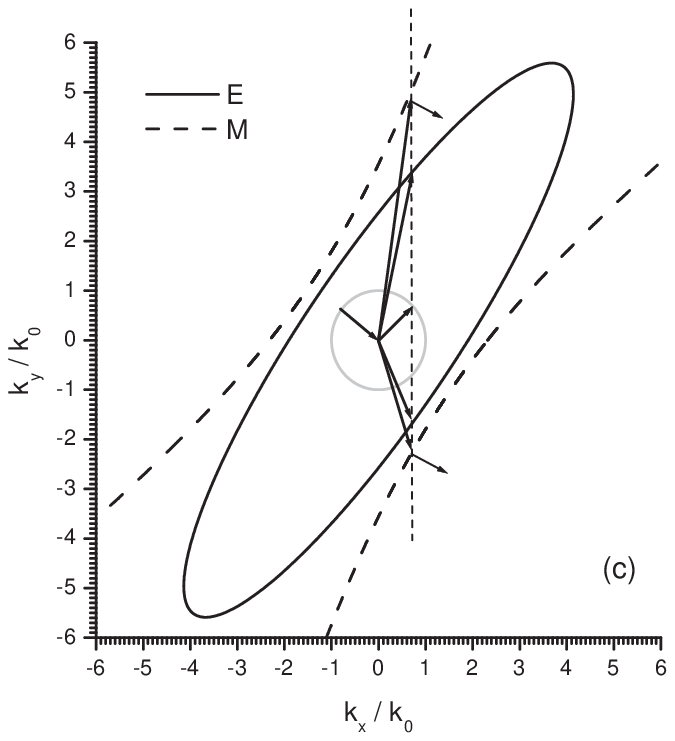} \hspace{0.3cm} 
\includegraphics[width=7.5cm]{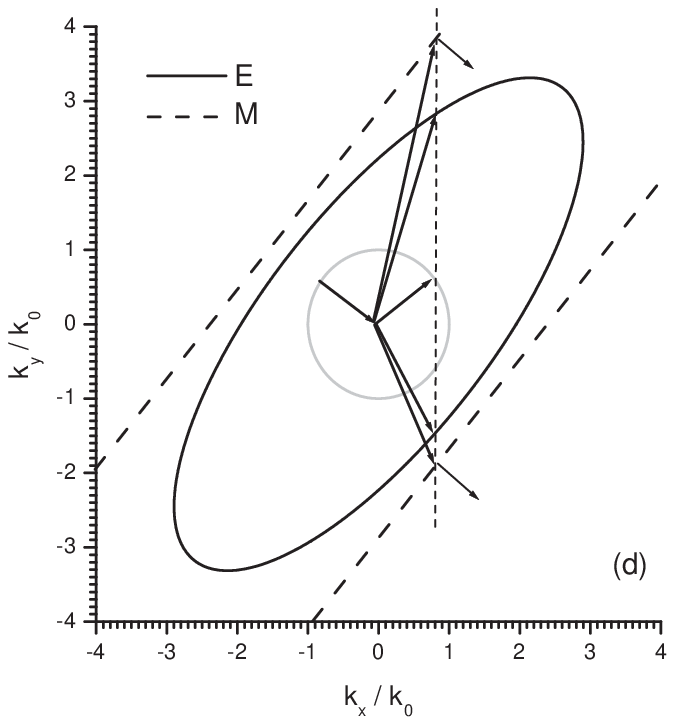} 
\end{tabular}
\end{center} 
\caption[example]{\label{phi60} 
Reciprocal space maps for Cases I and II, when $\varphi_c=60^\circ$. 
(a) $\theta_c=20^{\circ}$, 
(b) $\theta_c=52.73^{\circ}$, (c) $\theta_c=55^{\circ}$, and 
(d) $\theta_c=59.06^{\circ}$. The light gray circle represents the dispersion 
equation for plane waves in the medium of incidence. }
\end{figure}

In Figure \ref{phi60}, the reciprocal space maps for four 
different orientations of the optic axis are shown: 
\begin{itemize}
\item
$\theta_c=20^{\circ}$ (both dispersion curves hyperbolic), 
\item
$\theta_c=\theta_c^E=52.73^{\circ}$ (electric type linear and magnetic type 
hyperbolic), 
\item
$\theta_c=55^{\circ}$ (electric type elliptic and magnetic type 
hyperbolic), and 
\item 
$\theta_c=\theta_c^M=59.06^{\circ}$ (electric type elliptic
and magnetic type linear). 
\end{itemize}
For $\theta_c>\theta_c^M=59.06^{\circ}$, modes of both
electric and magnetic types have elliptic dispersion curves~---~just as 
for a natural 
crystal (not shown). The light gray circle in Figure \ref{phi60} 
represents the dispersion equation for plane waves in vacuum (the medium of incidence). 

For $\theta_c=20^\circ$, Figure \ref{phi60}a indicates
the nonexistence of real--valued $k_y$ in the refracting anisotropic medium 
for either the electric or the magnetic 
modes, the specific $k_x$ being indicated
by a dashed vertical line in the figure.
This is true for both  Cases I and II, for any angle of incidence
(with respect to the $y$ axis), and for any incident
polarization state; hence, the chosen anisotropic medium behaves
 as an omnidirectional total reflector \cite{hu-chui}. 
As the present--day construction of NPV metamaterials is such that the boundary 
is periodically stepped \cite{RDALDS04}, it is worth noting that the introduction 
of a periodic modulation along the surface would subvert the omnidirectional 
reflector effect, since a periodic modulation allows for the presence 
of spatial harmonics with tangential components of their wavevectors that 
can now satisfy the required matching condition. Gratings of this kind, 
contrary  to what happens for all gratings made of conventional materials, 
have been recently shown to support an infinite number of refracted channels 
\cite{GEM_NJP,GEM_scalar}. 

When $\theta_c=\theta_c^E=52.73^\circ$  
the dispersion equation for refracted modes of the electric type is linear. 
It is posible to find two wavevectors with real--valued
components that satisfy the phase--matching condition (the so--called Snell's
law)
at the interface, one belonging to the upper straight line and the other to the lower 
straight line in Figure \ref{phi60}b. 
As the direction of the time--averaged Poynting vector associated with electric 
modes is given by \cite{LVV91jmo}
\begin{equation}
\vec S =\frac{\omega\, \epsilon_{\perp}}{8 \pi\, \epsilon_{\parallel}}(\vec{k}\times \hat{c})^2\,\tilde{\epsilon}\cdot\vec{k}\;,\label{PoyntingE1}
\end{equation}
we conclude that the refracted wavevectors on the upper straight line do not satisfy 
the radiation condition for Case I, whereas wavevectors on the lower 
straight line do not satisfy the radiation condition for Case II. 

The direction of $\vec S$ given by (\ref{PoyntingE1}) for modes of the electric type is normal to the dispersion curves and points towards $y<0$, as required by the radiation condition. Ray directions coincide with the direction of $\vec S$. As for the parameters considered in our examples, the $z$ component of the time--averaged Poynting vector does not vanish, the ray directions are not contained in the plane of incidence. The projections of the refracted rays onto the $xy$ plane (indicated by little arrows in the figures) are perpendicular to the straight lines and independent of the angle of incidence. 

For refracted modes of the magnetic type and for the angle of incidence
($=\sin^{-1}k_x/k_o$)
 shown 
in Figure \ref{phi60}b, it is also posible to find two refracted 
wavevectors with real--valued
components satisfying the phase--matching condition at the interface, one belonging 
to the upper hyperbola (not shown) and the other to the lower hyperbola. 
The time--averaged Poynting vector associated with the magnetic modes is given by
\begin{equation}
\vec S =\frac{\omega}{8 \pi k_0^2}\frac{(\vec{k}\times \hat{c})^2}{\mu_{\perp}\mu_{\parallel}}\,\tilde{\mu}\cdot\vec{k}\;.\label{PoyntingM1}
\end{equation}
Therefore,
we conclude that wavevectors on the upper hyperbola do not satisfy 
the radiation condition for Case II, whereas wavevectors on the lower 
hyperbola do not satisfy the radiation condition for Case I. 
Ray directions  coincide with the direction of $\vec S$ given by \ref{PoyntingM1}, which again has a non--zero component in the $z$ direction. Ray projections onto the $xy$ plane (indicated by little arrows in the figures) are perpendicular to the hyperbolas.
 
The interface for both Cases I and II acts as a positively refracting interface for modes of both types, in the sense that the refracted rays never emerge on the same side of the normal as 
the incident ray \cite{counterposedLM}.

When the angle $\theta_c$ is increased  to $55^\circ$ (Figure \ref{phi60}c), the dispersion equation for the refracted modes of the magnetic type is still hyperbolic, but the dispersion equation for the electric type is elliptic.
Again, for both electric and magnetic modes, is it possible to find two wavevectors with 
acceptable real--valued components. From (\ref{PoyntingE1}), we conclude that
refracted electric modes on the upper part of the ellipse correspond to Case II, whereas electric wavevectors on the lower part of the ellipse correspond to Case I. On the other hand, wavevectors for the refracted magnetic modes on the upper hyperbola do not satisfy the radiation condition for Case II, whereas wavevectors on the lower hyperbola do not satisfy the radiation condition for Case I, as can be deduced from (\ref{PoyntingM1}). 

Ray projections onto the $xy$ plane corresponding to the magnetic modes alone
are shown in the figure, for the sake of clarity. 
For both Cases I and II and for refracted modes of the electric and magnetic types, the refracted rays never emerge on the same side of the $y$ axis as the incident ray, just as for positively refracting interfaces.

When $\theta_c=\theta_c^M=59.06^\circ$ (Figure \ref{phi60}d), the dispersion
curves for the refracted modes of the electric type continue to be ellipses, but now the 
dispersion curves   for the modes of the magnetic type become straight lines.
For the electric modes, the selection of the wavevectors is identical to that in
Figure \ref{phi60}c. For the refracted
magnetic modes, wavevectors on the upper straight line do not satisfy 
the radiation condition for Case II, whereas wavevectors on the lower 
straight line do not satisfy the radiation condition for Case I. 

Ray projections onto the $xy$ plane for the refracted magnetic modes  are also drawn in the figure.
Again, for both Cases I and II the surface acts as a positively refracting interface 
for modes of both types.

\section{Concluding remarks}\label{sec:conc}

This work focused on the geometric representation at a fixed frequency of the
dispersion surface $\omega(\vec k)$ for lossless, homogeneous  
dielectric--magnetic uniaxial materials. To encompass both natural crystals 
and the artificial composites used to demonstrate negative refraction (metamaterials), 
we assumed that the elements of the permittivity and permeability tensors 
characterizing the material can have any sign. We showed that,
depending on a particular combination of the elements of these tensors, 
the propagating electromagnetic modes supported by 
the material can exhibit dispersion surfaces in the form of (a) ellipsoids of revolution, 
(b) hyperboloids of one sheet, or (c) hyperboloids of two sheets. Intersections of these surfaces with  fixed planes of propagation lead to circles, ellipses, hyperbolas or straight lines, depending on the relative orientation of the optic axis. 
This analysis was used to discuss the reflection and refraction of 
electromagnetic plane waves due to a planar interface with vacuum (or any linear,
homogeneous, isotropic, dielectric--magnetic medium).


\vskip 0.5cm

\noindent {\bf Acknowledgments}  RAD and MEI acknowledge financial support from Consejo Nacional de Investigaciones Cient\'{\i}ficas y T\'{e}cnicas (CONICET), Agencia Nacional de Promoci\'{o}n Cient\'{\i}fica y Tecnol\'{o}gica (ANPCYT-BID 1201/OC-AR-PICT14099) and Universidad de Buenos Aires. AL is grateful for financial support from the Penn State CIRTL project.

\newpage

\baselineskip 0.5 cm
\begin{center}
Table 1: Types of possible dispersion surfaces for different combinations between 
the eigenvalues $\epsilon_{\perp}$, $\epsilon_{\parallel}$, 
$\mu_{\perp}$ and $\mu_{\parallel}$ of the real symmetric tensors 
$\tilde \epsilon$ and $\tilde \mu$. The first symbol indicates the mode: E (electric) 
or M (magnetic). The second symbol indicates the geometrical form of 
the dispersion surface: $e$ (ellipsoids of revolution), $h_1$ (hyperboloid of one sheet), 
$h_2$ (hyperboloid of two sheets). The symbol $n$ indicates that the
corresponding mode is of the evanescent (i.e., nonpropagating) kind. \\ 
\vspace{0.8cm}
\begin{tabular}{|c|c|c|c|c|}
\hline
$\,$  & $\epsilon_\perp>0$ & $\epsilon_\perp>0$ & $\epsilon_\perp<0$ & $\epsilon_\perp<0$ \\
$\,$  & $\epsilon_\parallel>0$ & $\epsilon_\parallel<0$ & $\epsilon_\parallel>0$ & $\epsilon_\parallel<0$ \\
\hline
$\mu_\perp>0$        & E\,$e$   &  E\,$h_2$  & E\,$h_1$ & E\,$n$     \\
$\mu_\parallel>0$    & M\,$e$   &  M\,$e$    & M\,$n$   & M\,$n$     \\
\hline
$\mu_\perp>0$        & E\,$e$   &  E\,$h_2$ & E\,$h_1$ & E\,$n$     \\
$\mu_\parallel<0$    & M\,$h_2$ &  M\,$h_2$ & M\,$h_1$ & M\,$h_1$   \\
\hline
$\mu_\perp<0$        & E\,$n$   &  E\,$h_1$ & E\,$h_2$ & E\,$e$     \\
$\mu_\parallel>0$    & M\,$h_1$ &  M\,$h_1$ & M\,$h_2$ & M\,$h_2$   \\
\hline
$\mu_\perp<0$        & E\,$n$   &  E\,$h_1$ & E\,$h_2$ & E\,$e$     \\
$\mu_\parallel<0$    & M\,$n$   &  M\,$n$   & M\,$e$   & M\,$e$     \\
\hline
\end{tabular}
\end{center}
\vspace{0.8cm}
\newpage

\newpage

\end{document}